\documentclass[final,5p,twocolumn]{elsarticle}

\usepackage[english]{babel}
\usepackage[utf8x]{inputenc}
\usepackage{amsmath}
\usepackage{graphicx}

\usepackage{url}

\usepackage{caption}
\usepackage{subcaption}
\usepackage{tabularx}

\usepackage{listings}

\usepackage{booktabs}

\journal{---}
\begin{document}

\begin{frontmatter}

\title{Individuals, Institutions, and Innovation in the  \\ Debates of the French Revolution}

\author[1]{Alexander T. J. Barron}
\author[1,4]{Jenny Huang} 
\author[2]{Rebecca L. Spang}
\author[3,4]{Simon DeDeo\corref{cor1}}
\ead{sdedeo@andrew.cmu.edu}
\cortext[cor1]{To whom correspondence should be addressed.}

\address[1]{\footnotesize School of Informatics, Computing, and Engineering, Indiana University, 919 E 10th Street, Bloomington, IN 47408, USA}
\address[2]{\footnotesize Department of History, Indiana University, 1020 E Kirkwood Ave, Bloomington, IN 47405, USA}
\address[3]{\footnotesize Department of Social and Decision Sciences, Carnegie Mellon University, 5000 Forbes Avenue, BP 208, Pittsburgh, PA 15213, USA}
\address[4]{\footnotesize Santa Fe Institute, 1399 Hyde Park Road, Santa Fe, NM 87501, USA}

\begin{abstract}
The French Revolution brought principles of ``liberty, equality, and brotherhood'' to bear on the day-to-day challenges of governing what was then the largest country in Europe. Its experiments provided a model for future revolutions and democracies across the globe, but this first modern revolution had no model to follow. Using reconstructed transcripts of debates held in the Revolution's first parliament, we present a quantitative analysis of how this system managed innovation. We use information theory to track the creation, transmission, and destruction of patterns of word-use across over 40,000 speeches and more than one thousand speakers. The parliament as a whole was biased toward the adoption of new patterns, but speakers' individual qualities could break these overall trends. Speakers on the left innovated at higher rates while speakers on the right acted, often successfully, to preserve prior patterns. Key players such as Robespierre (on the left) and Abb\'{e} Maury (on the right) played information-processing roles emblematic of their politics. Newly-created organizational functions---such as the Assembly's President and committee chairs---had significant effects on debate outcomes, and a distinct transition appears mid-way through the parliament when committees, external to the debate process, gain new powers to ``propose and dispose'' to the body as a whole. Taken together, these quantitative results align with existing qualitative interpretations but also reveal crucial information-processing dynamics that have hitherto been overlooked. Great orators had the public's attention, but deputies (mostly on the political left) who mastered the committee system gained new powers to shape revolutionary legislation. 
\end{abstract}


\end{frontmatter}






\clearpage
\noindent
The French Revolution was a turning point in European history. Revolutionary commitments to individual liberty collided with ideals of social equality, while the rejection of Divine-Right monarchy and the embrace of laws based on reason opened a host of practical questions about how to govern the most populous state in Europe. The first parliament of the Revolution, the National Constituent Assembly (NCA), was, not surprisingly, a picture of upheaval from its outset.

Over the course of two years, the thousand or more individuals in that Assembly took it upon themselves to propose and argue the previously unimaginable: the revocation of Old-Regime privilege and the reinvention of the relationship between individual and state. But this parliament was more than a debate society for ambitious young men. It was also the origin of a system of rule. In the years that followed, successive legislative bodies declared war on most of Europe, dissolved the French monarchy, declared a Republic, and sentenced the former king to death---all while simultaneously writing constitutions and passing ordinary legislation. Many of their procedures, and some of their personnel, were drawn from the experience of the NCA.

As a parliament, the NCA faced the problems that come with managing massive flows of information from the outside---problems that face the modern deliberative political bodies that, in many cases, are the NCA's direct descendants~\cite{jones2005politics}. But as the \emph{first} parliament they also shared the challenges of the knowledge-seekers of the Enlightenment and their descendants: speakers faced the risks and rewards of innovation as they introduced ideas and attempted, under unusual conditions, to achieve lasting influence over the future arguments of their contemporaries~\cite{evans}.

Key questions arise from seeing the NCA as simultaneously a site of political and epistemic activity. On the epistemic side: How did new ideas enter that parliament-room, and how were they taken up, or discarded, by those who heard them? What kinds of roles did individuals play in this process? On the political side: What institutions did the parliament evolve to manage the onslaught of novelty and reaction, optimism and grievance, philosophical argument and organizational minutiae that characterized the day-to-day tasks of governance and nation-building?

The digitization of historical archives allows us to answer these questions in a fundamentally new way. Using latent Dirichlet allocation~\cite{Blei:2003tn} and new techniques in information theory drawn from the cognitive sciences~\cite{Murdock2017117}, we track the emergence and persistence of word-usage patterns in over 40,000 speeches made in the NCA and later reconstructed in the \emph{Archives Parliamentaires} (AP) from detailed records kept at the time of the Revolution. Two critical measures, novelty (how unexpected a speech's patterns are, given past speeches) and transience (the extent to which those patterns fade away, or, conversely, persist in future speeches), allow us to trace both emergent political institutions and the generation and propagation of new ideas and manners of speech. Viewing the turbulent early days of the French Revolution through the lens of pattern creation, sharing, and destruction provides a complementary viewpoint to the study of specific ideas and particular historical events. This distinction can also be found in evolutionary biology, where one can study either (on the one hand) the mechanisms of transmission and selection or (on the other) the particular phenotypes for which an environment selects.

We find, at high significance, a bias in favor of the propagation of novel patterns from prior speeches into the future: in the framework of cultural evolution, the flow of ideas through NCA is out of equilibrium and the system preferentially selects for what violates prior expectations. This effect is driven in part by charismatic political radicals such as Robespierre and P\'{e}tion de Villeneuve, who not only introduced new patterns at higher rates than their peers, but managed to do so in a way that led others to copy those patterns forward. By contrast, influential conservative figures such as Abb\'{e} Maury and Cazal\`{e}s acted as inertial dampeners, successfully preserving prior patterns in the face of an overall revolutionary bias towards innovation. Conservatives ``conserve'': not just by referring to past traditions, but also in the day-to-day dynamics of discussion, where they act to keep conversations on track.

In parallel with these individual-level differences, our methods reveal a major transition in how the parliament as a whole processed novelty, occurring about half-way through the parliament's lifetime. This shift is associated with committees that reported to, but met outside of, the parliament itself, and that had gained new power both to raise and resolve questions. While orators on the left and right continued to confront each other in public speeches from the floor, it was the left that captured this new institutional mechanism and used it to their own advantage. The consolidation of this structural shift was accompanied by radicalization of the left and an accelerating flight of conservatives from both the parliament and the country itself.


\section*{Results}

Social systems are characterized by heteroglossia: the coexistence, sharing, and competition of different patterns of speech.  Heteroglossia makes linguistics and rhetoric, both concerned with the study of the reception, influence, and propagation of language within a community~\cite{benoit2003rhetorical,harris2005reception}, core components in the quantitative study of culture. Tracking changes in speech patterns within a social body allows us to examine cultural evolution: the circulation, selection, and differential propagation of speech patterns in the group as a whole (rather than, say, tracking the ideas of a single individual). Patterns of heteroglossia demonstrate existing power relations, create new ones, and are a key method for the definition of both institutions and genres~\cite{foucault1980power,danescu2012echoes,Klingenstein01072014}. Our methods here quantify a key aspect of cultural evolution: the extent to which one agent's language patterns are used and copied by another~\cite{bakhtin2010dialogic,blythe2012s}. 

To study the flow of rhetorical influence and attention in the NCA over time, we characterize how patterns of language use, uncovered by topic modeling, are propagated from speech to speech. We do so using Kullback-Leibler Divergence~\citep[KLD]{Kullback1951}: KLD, or ``surprise'', from one pattern to another measures the extent to which the expectations of an optimal learner, trained on the first pattern, are violated by the second. Other work has demonstrated that surprise (in the Kullback-Leibler sense of the word) is a cognitive as well as an information-theoretic quantity. It predicts, for example, what a subject will look at in a dynamically-evolving visual scene~\cite{itti2009bayesian}. KLD can be used to map an individual's higher-level activities as well, detecting, for example, biographically-significant transitions in a subject's intellectual life~\cite{Murdock2017117}.

Here, we use surprise to analyze a corpus of speeches by many different individuals. Surprise then measures both the extent to which a particular speech deviates from the patterns of prior speeches (novelty), and from the patterns that will appear in the future (transience). High surprise compared to the past indicates the topic mixture is new compared to previous speeches, hence the term ``novelty''; high surprise compared to the future indicates that later speeches do not retain that pattern very strongly, hence the term ``transience''. 

\subsection*{Innovation Bias}

\begin{figure*}
\centering
\includegraphics[width=\textwidth]{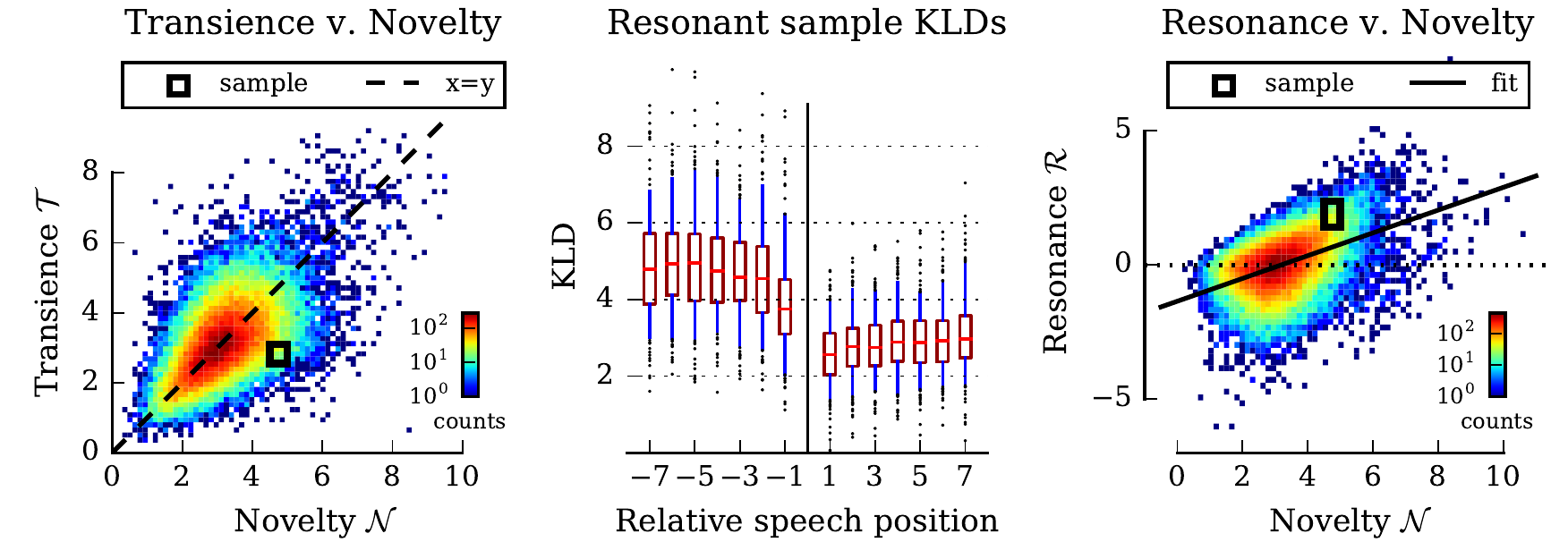}
\caption{Novelty, transience, and resonance in the French Revolution. Left: a density plot of transience vs. novelty at scale $w=7$.  Speeches close to the identity ($x=y$) line have equal surprise from the past and to the future.  Resonant speeches, whose transience is anomalously low compared to their novelty, break this symmetry and fall below the identity line; the middle plot shows this asymmetry for a marked sample of high-resonance speeches. Right: resonance vs. novelty, with regression line.  Although novelty is tied to transience, and therefore risk of irrelevance, it is also directly proportional to resonance.}
\label{fig:NTR_pedagogical}
\end{figure*}

Speeches in the NCA span a wide range in both novelty and transience; Fig.~\ref{fig:NTR_pedagogical} summarizes the system at the level of individual speeches, in this case at the relatively rapid timescale (window width, $w$) of seven speeches. While the majority of speeches concentrate near the symmetry line---speeches with high novelty are likely to have similarly high transience---two results stand out. First, the scatter is large: there are many speeches that lie far off the novelty--transience line of equality, and it is easy (for example) to find speeches with top-quartile novelty that have bottom-quartile transience. ``What is new is quickly forgotten'' is a useful heuristic, but holds only in the average. Below, we consider two potential drivers of this diversity of reception: the speaker, and the context in which the speech was made.

Second, novel speeches are unexpectedly influential. We quantify this with ``resonance,'' the imbalance between novelty and transience (see Materials and Methods). Resonance, the quality of at once breaking from the past and influencing the future, increases with novelty, as shown in the rightmost plot of Fig.~\ref{fig:NTR_pedagogical}.  We refer to this positive relationship as innovation bias: penalties to high novelty speeches are lower than expected in a system at equilibrium. This bias is measured by $\Gamma$, the slope of the novelty--resonance line; positive $\Gamma$ indicates innovation bias. We find this bias in place from the most rapid timescales (one speech to the next, $w=1$), all the way up to $w\approx100$, timescales equivalent to a day or so (see SI). This innovation bias lasts for at least the course of a day, as speakers deliberately turn to new topics, but fades away on longer timescales.

\subsection*{Organizational roles and individual strategies}

In choosing when to speak and what to say, speakers had some control over the relative novelty of their speeches. A speaker's attitude toward the Assembly as an institution, as well as other political or philosophical commitments, would contribute to his willingness to create new word patterns or copy those of earlier speakers. Conversely, speakers had much less control over the reception of what they said. Idiosyncratic properties of the speaker, ranging from the political (\emph{e.g.}, party membership) and the social (\emph{e.g.}, demeanor or prestige) to the rhetorical (everything from word choice to pitch and volume of speech) would have altered the reception of their words, raising or lowering the transience of the patterns they create.

Differences in strategy and reception at the individual level, in other words, could break the system-level trends in favor of innovation. An unskilled but risk-tolerant delegate might have succeeded only rarely, so that his speeches, while novel, tended to have low, or even negative, resonance. Conversely, the prestige or social power of a speaker who wished to maintain a line of argument might have allowed him to simultaneously maintain low novelty and high resonance, effectively keeping the conversation on track.

Individual strategies were not the only source of deviation from system-level innovation bias. From the first days of its meeting, the Assembly organized itself in such a way as to assign explicit roles to particular speakers. These information-processing functions over-ruled an individual's personal characteristics. An example in the NCA, and common to many later parliaments today, is the role of president, who served as both a point of contact for the King and an enforcer of the daily agenda~\cite{Tackett:2014us}. The president was tasked with a code of conduct and functional role, and over the course of the NCA 49 delegates served in this position.

The NCA also created another specialized entity: the committee. Committees, whose members were notionally selected on the basis of relevant expertise, deliberated in private. They developed content outside the debate process and then presented it to the full body for public review. While a speaker on the floor of the Assembly might play to the audience in the visitor's galleries, committee members addressed only each other.

Lexical markers in our data identify when a committee proxy was speaking, and allow us to classify their speech into two  categories: ``new-item'' speech and ``in-debate'' committee speech (see SI). New-item speeches introduced official content to the Assembly floor, typically articles and draft decrees to be reviewed by the legislative body, and mark transitions in attention from one topic to another. In-debate speeches occurred when a committee member or reporter engaged with other delegates following the introduction of the item.

To understand how both individual-level strategies and system-imposed roles affected the production and reception of new patterns, we consider the novelty and resonance of speeches given by those holding three distinct roles in the Assembly: the 40 most common orators, the President (regardless of who held the position at the time), and committee proxies.  We calculate both the average novelty and resonance for each category (scaled by $z$-score), $z(\mathcal{N})$ and $z(\mathcal{R})$, tracking both potentially idiosyncratic novelty-seeking (or avoidant) strategies, and how they are received by the system as a whole.  We do this at scales of $w$ from one (one speech compared to either the next, or prior, speech) to 5,000 (speech compared to, roughly, three months of speeches before or after).

The system's overall bias in favor of innovation would predict that the category of speakers with the highest novelty would also have the highest resonance. To determine if this expectation is validated in our sources, we compare the overall novelty-resonance relationship of the system (depicted by the fit line in Fig.~\ref{fig:NTR_pedagogical}) to the measured resonance for speeches from each of the three categories (orator, President, committee proxy). Specifically, for each category we report $\Delta z(R)$, defined as $z(\mathcal{R}) - \mathbf{E} [ z(\mathcal{R}) | z(\mathcal{N}) ]$, the difference between the measured mean resonance of speeches and the expected mean resonance under the OLS model $z(\mathcal{R}) \sim z(\mathcal{N})$.  While $z(\mathcal{R})$ measures the effect speakers had on later discourse, $\Delta z(\mathcal{R})$ measures the extent to which they broke system-level trends in achieving those effects. For example, a speaker with high novelty may have high resonance but negative $\Delta z(\mathcal{R})$, indicating that his adventurousness was rewarded less than expected. Full results for $w=36$ (roughly half a day) are shown in Table~\ref{tab:entities_categorized}; results for novelty are stable on all timescales, while resonance at $w=36$ is strongly correlated from $w=3$ to $w=100$ (see SI). 
\begin{table}[h!]
\centering
\resizebox{0.5\textwidth}{!}{
\begin{tabular}{lllll}
\toprule
                         Name & $z(\mathcal{N})$ &      $z(\mathcal{R})$  & $\Delta z(\mathcal{R})$ & Type  \\
\midrule
\multicolumn{5}{c}{High novelty, high resonance}\\
\midrule
J\'{e}r\^{o}me P\'{e}tion de Villeneuve & 0.10 & 0.28*** & +0.25*** & 3g \\
Maximilien Robespierre & 0.11** & 0.18*** & +0.14** & 3g \\
Jean-Denis Lanjuinais & 0.06 & 0.16*** & +0.15*** & 3g \\
Alexandre Lameth & 0.17** & 0.14* & +0.09 & 2g \\
Charles Antoine Chasset & 0.31*** & 0.13* & +0.04 & 3g \\
\textbf{Committee (new item)} & 1.31*** & 0.12*** & -0.27*** & -- \\
Philippe-Antoine Merlin & 0.27*** & 0.05 & -0.03 & 3g \\
Pierre-Fran\c{c}ois Gossin & 0.65*** & 0.03 & -0.17** & 3g \\
Jacques Fran\c{c}ois Menou & 0.40*** & 0.02 & -0.10 & 2g \\
\textbf{Committee (in debate)} & 0.29*** & 0.02 & -0.07*** & -- \\
\textbf{Left Wing} & 0.07*** & 0.02* & 0.00 & (g) \\
\textbf{3rd Estate} & 0.06*** & 0.01 & -0.02* & -- \\ 
\midrule
\multicolumn{5}{c}{High novelty, low resonance}\\
\midrule

Jacques Guillaume Thouret & 0.16*** & 0.00 & -0.05 & 3g \\
Jac. Jo. Def. de Chapeli\'{e}res & 0.35*** & -0.03 & -0.13** & 3- \\
Fran\c{c}ois Denis Tronchet & 0.24*** & -0.04 & -0.11 & 3g \\
Armand-Gaston Camus & 0.29*** & -0.04 & -0.13*** & 3g \\
\textbf{President} & 0.02 & -0.07*** & -0.08*** & -- \\
Th\'{e}odore Vernier & 0.55*** & -0.14** & -0.31*** & 3g \\

\midrule
\multicolumn{5}{c}{Low novelty, high resonance}\\
\midrule

Gu. Fr. Ch. Goupil-Pr\'{e}felne & -0.21*** & 0.13** & +0.20*** & 3g \\
Jean-Fran\c{c}ois Reubell & -0.18*** & 0.11* & +0.16*** & 3g \\
Louis Simon Martineau & -0.05 & 0.08* & +0.10* & 3g \\
Jacques An. Ma. de Cazal\`{e}s & -0.44*** & 0.08* & +0.21*** & 2d \\
Pierre Victor Malouet & -0.27*** & 0.08* & +0.16*** & 3d \\
Jean-Sifrein Maury  & -0.46*** & 0.07 & +0.20*** & 1d \\
Pierre-Louis Prieur & -0.27*** & 0.05 & +0.13** & 3g \\
\textbf{1st \& 2nd Estates} & -0.10*** & 0.03*** & +0.05*** & -- \\
Jean-Fr. Gaultier de Biauzat & -0.13** & 0.03 & +0.06 & 3g \\
\textbf{Right Wing} & -0.32*** & 0.03* & +0.10*** & (d) \\
\midrule
\multicolumn{5}{c}{Low novelty, low resonance}\\
\midrule

Dominique Garat & -0.13** & 0.00 & +0.04 & 3g \\
Antoine Ch. Ga. Folleville & -0.44*** & -0.01 & +0.12* & 2d \\
Lo.-Mi. Le Pel. de Saint-Fargeau & -0.20*** & -0.01 & +0.05 & 2g \\
Fr. Do. de Reynaud Montlosier & -0.61*** & -0.02 & +0.17* & 2d \\
Pierre-Louis Roederer & -0.10* & -0.03 & 0.0 & 3g \\
Louis Foucauld de Lardimalie & -0.53*** & -0.05 & +0.11 & 2d \\
Charles Malo Lameth & -0.15*** & -0.06 & -0.02 & 2g \\
Pierre Fran\c{}ois Bouche & -0.09** & -0.10* & -0.07* & 3g \\
Antoine Barnave & -0.04 & -0.12* & -0.11* & 3g \\
\end{tabular}}
\caption{Mean novelty and resonance per speaker at scale 36, for role and type (in bold) and the top forty orators. $z(\mathcal{N})$: novelty compared to system average; $z(\mathcal{R})$: resonance compared to system average; $\Delta z(\mathcal{R})$: resonance relative to predicted resonance given novelty.  ``Type'' codes for estate (3: bourgeoisie; 2: nobility; 1: clergy) and political affiliation (g: \emph{gauche}, left-; d: \emph{droit}, right-wing).} 
\label{tab:entities_categorized}
\end{table}

Institutional roles---speaking on behalf of a committee, presiding over the Assembly---did not follow these system-level trends. High novelty and high resonance together characterize committees as gatherers of new information that they injected into debate in ways that defined downstream discussion. In contrast, the president's role as agenda-enforcer led to lower than average resonance: he acted, at best, to summarize what had come before, while having less influence on the patterns of speech that followed.  Though he might break from conversation to further the agenda, the content he introduced to do this tended not to persist. The overall novelty-bias can not be explained by the taking-in of new information from committees: while committees did achieve above-average resonance ($z(\mathcal{R})$ greater than zero), they achieved lower resonance than expected ($\Delta z(\mathcal{R})$ less than zero). 

Individual strategies are often as defining as top-down roles. Of the top forty speakers in the assembly, thirty show significant deviations from aggregate patterns in either novelty or resonance at the $p<0.05$ level, with twenty-three speakers showing deviations at $p<10^{-3}$. Speakers deviate in both directions, with some showing anomalously high tendencies to break with past patterns, and others showing similarly strong tendencies to preserve them. High-novelty speakers are overwhelmingly associated with the left wing and the bourgeoisie, while all of our right wing speakers, and the vast majority of nobility, are low-novelty.

More than half the speakers in our data pursued strategies that distinguished them from the aggregate; significant differences in resonance, by contrast, are less common. The speakers that \emph{do} show differences in reception, however, were among the key players of the Revolution. The celebrated radicals Robespierre and P\'{e}tion not only achieved the highest average resonance in the system, but significantly higher resonance than even that due to the system-wide novelty bias (positive $\Delta z(\mathcal{R})$). Conversely, speakers such as Armand-Gaston Camus and Th\'{e}odore Vernier, called on primarily for their specialized knowledge in canon law and taxation, respectively, show high-novelty, but low resonance: they presented information, but either lacked the lasting influence of  more prominent speakers, or were able to settle questions so conclusively that the room moved on. Finally, prominent political conservatives such as Jean-Sifrein Maury and Jacques de Cazal\`{e}s appear in the low-novelty, high-resonance quadrant. They were able to break the system-level novelty bias, and are notable not only for keeping the conversation on track (low novelty), but for being able to influence others to do the same (high resonance).  In this, Maury and Cazal\'{e}s are characteristic of the right-wing overall: while the novelty-biased left was composed of both high and low resonance speakers, the right-wing was able to achieve system influence (positive $z(\mathcal{R})$ and $\Delta z(\mathcal{R})$) despite their anomalously low novelty.

\subsection*{The Emergence and Evolution of the Committee}

Committees were a key innovation of the NCA, and allowed the system to manage large amounts of information without overwhelming the discussions in the parliament-room itself. They did not, however, appear overnight. While the previous section establishes their unusual functional role, the comprehensive coverage of the AP makes it possible to study how their role emerged. In this section, we show how returns to novelty, $\Gamma$, are modulated by committee roles over time. We fit, separately, two terms that quantify the additional boost (or decrement) to the novelty-resonance relationship when speeches either introduce new committee items ($\Gamma_n$), or advocate on behalf of a committee during debate ($\Gamma_d$). A speech of novelty $\mathcal{N}$, for example, achieves on average a resonance $\mathcal{R}$ equal to $(\Gamma+\Gamma_n)(\mathcal{N}-\mathcal{N}_0)$ when made by a committee member introducing a new item, compared to $\Gamma\mathcal(\mathcal{N}-\mathcal{N}_0)$ when the speaker acts on his own behalf.

We look for discrete shifts in committee function, doing change-point detection with a maximum-likelihood model of the novelty-resonance relationship where $\Gamma$, $\Gamma_n$, and $\Gamma_d$ are allowed to vary in time. Following Ref.~\cite{Murdock2017117}, we consider a two-epoch model, where all three quantities are fixed to constant values in each epoch, with a single discrete change at a particular time-point whose position is a free parameter. The two-epoch model is preferred to a single-epoch model, as well as to a linear (secular shift) model under AIC. We find strong evidence for a change-point in the nature of committee functions occurring at the end of 1790; the modal best-fit date across all scales is 31 October 1790. Allowing the intercepts of the new-item and in-debate speeches, as well as their slopes, to vary produces nearly identical results (see SI).\begin{figure*}
\centering
\includegraphics[width=0.5\textwidth]{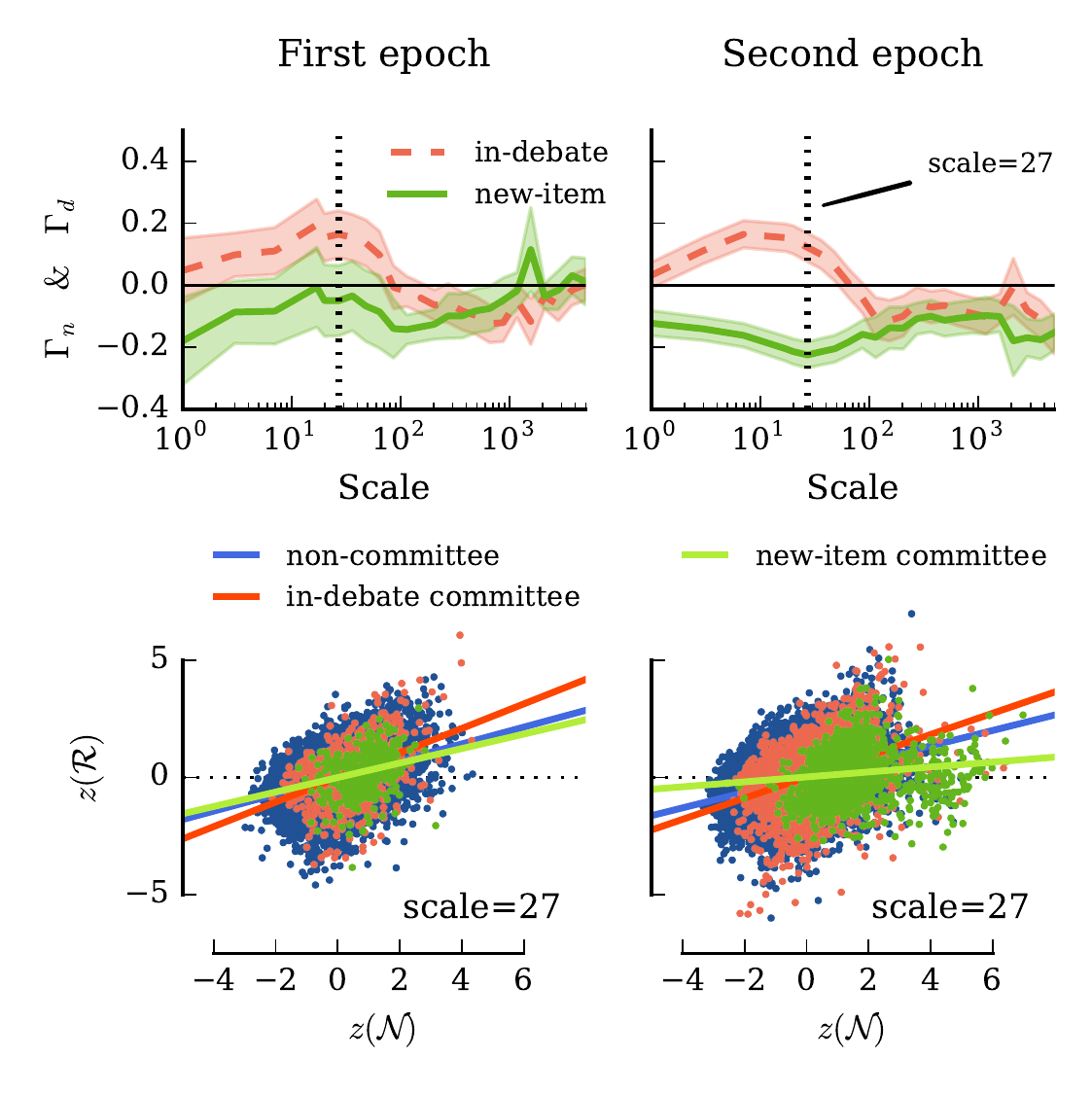} 
\caption{Information-processing functions of NCA committees before (1st column) and after (2nd column) the late-1790 change-point. Top row: the shift in the novelty-resonance relationship for new-item and in-debate committee speech, with 99\% confidence intervals.  Bottom row: scatter plots and fit lines at scale 27 for these speech types, compared to all other speeches. The ``undebated tail'' is seen in the second epoch, as committees gain new powers to ``propose and dispose''.}
\label{fig:datafit_histinsp}
\end{figure*}
Fig.~\ref{fig:datafit_histinsp} displays the novelty-transience relationships for the two epochs, demonstrating the emergence of a new role for committee speech over time. In the first epoch, the resonance of new items was indistinguishable from speech of similar novelty in the system as a whole. However, speech received a resonance bonus when delivered by a committee member as part of a debate. In other words, committee representatives injected new information in a fashion similar to other delegates (new-item speech), but had privileged abilities in guiding the subsequent debate (in-debate speech). 

The pattern is different in the second epoch, where new-item speech has anomalously low resonance at high novelty. Inspection of the speeches themselves suggests that this high-novelty/low-resonance tail is associated not with (as in the case of individual speakers) failure to alter the course of debate, but rather with the increasing power committees had to ``propose and dispose'': once the committee presented their findings to the parliament, they were increasingly accepted with minimal discussion. In some cases, committee reports are noted as ``passed without debate''; for a sample of these cases, we find the mean $z(\mathcal{N})$ equal to $1.57\pm0.16$ and $z(\mathcal{R})$ equal to $0.14\pm0.13$, \emph{i.e.}, committee speeches with these hand-annotated outcomes follow a similarly high-novelty, low-resonance pattern to other new-item speeches, further validating the interpretation of the ``undebated tail'' as a signal of emergent committee power. This power can also be seen when substantive debate on committee matters does occur: ``in debate'' speeches made by committee members retain a privileged role in fixing and propagating the patterns that define that debate. 

\section*{Discussion}

The turbulent months at the beginning of the French Revolution led to a durable transformation in the very nature of European government~\cite{sewell1996historical}. Our study considers the elite debates at the center of these events, focusing on pattern transmission rather than the differential propagation of the ideas these patterns may communicate. This provides a new window into qualitative analyses, which usually focus on analyzing the logic of particular ideas and arguments over time~\cite{baker1990inventing, baker2015scripting}. 

Our analysis reveals clear differences between how the left- and right-wing figures created and transmitted patterns of language use. Conservatives may indeed ``stand athwart history, yelling Stop''~\cite{buckley1955our}; our results show that here they did so less by redirecting conversation from one set of patterns to another than by maintaining the already-established patterns of the conversations in which they participated. Indeed, the spatial metaphor of ``left vs. right'' is itself misleading: from the point of view of the debates themselves, the right wing acts as an inertial center, holding off conversation drift, while left-wing speakers produce a wide spectrum of innovations on a much larger periphery, only some of which survive. These roles become visible without reference to the ideas that are propagated: Robespierre emerges as a high-novelty, high-resonance speaker even before we consider the content of the speeches that made him an icon of revolutionary politics.

We also find an emergent distinction between orators and institution players. Talented speakers like Robespierre, P\'{e}tion, and Maury could achieve power through rhetoric and debate, appealing to the crowds in the galleries as much as their colleagues. Yet not every ambitious delegate could take this road. Quite apart from a delegate's rhetorical capacities or political stature, the physical venue itself was demanding: loud and (literally) resonant voices were required to be heard. While the NCA provided a place for rhetorical masters to thrive, in other words, it also ended up producing a second class of delegate who concentrated on committee work. Deliberating outside the parliament-room, committees contributed specialized knowledge and abilities and became indispensable workhorses for the emerging government~\cite{Tackett:2014us}. The information-processing functions that committees took on distinguish them clearly from other forms of speech, while Fig.~\ref{fig:datafit_histinsp} shows how these functions emerged over time; once in place, committees are both strong sources of new patterns of speech and new sources of extra-rhetorical power. Both individuals and institutions, in other words, mattered, playing distinct roles in the development of the parliament over time.

Many delegates welcomed these extra-parliamentary functions: committees, wrote the assembly member Jacques Dinochau, ``regulate the order of debate, classify questions, and maintain a continuity of principles, thus preventing an incoherence which might otherwise have menaced our decrees'' ~\cite{Tackett:2014us}. Yet the private nature of these committees was in dramatic contrast to the public debates in the hall itself; their increasing influence a testimony to the emergent distinction between the spectacle of democracy and the actual ways functioned in a body and polity too large for direct participation. The dramatic and early appearance of a specialized information-processing role for committees provides a new view on their place in modern democracies, where they emerge endogenously~\cite{inform_committee} to serve as essential information-management systems~\cite{survey_comm}.


The power of these committees continued to grow. While early committees were devoted to technical matters such as monetary and fiscal policy, they were also instrumental at key moments such as the dismantlement of feudal and religious privileges. When the revolution collapsed into chaos in 1793, it was committees, such as the famous ``Committee of Public Safety'' with Robespierre at its head, that effectively replaced the republican government and executed thousands of ``enemies of the revolution'' including many of the speakers in Table~\ref{tab:entities_categorized}. 

\section*{Conclusion}

The history of human culture is more than just the rise and fall of particular ideas. It is also the emergence of new information-processing mechanisms and media, and roles that individuals and institutions could play in creating and selectively propagating these ideas through time. In the language of biological evolution, we must understand not only the particular characteristics for which an environment selects, but the strength of that selection over time, and the shifting and heterogenous nature of the transmission mechanisms. 

By quantifying the flow of patterns between speakers, we can see revolutionary debates as not just a battle of ideas, but as struggle for the direction of the conversation itself. At the earliest stages of the revolution, political players adopted not just different ideological positions, but different roles in the propagation of patterns, with varying levels of innovation and fidelity to the past. And together, both cooperatively and agonistically, they invented new mechanisms for the collective management of information. 


\section*{Materials and Methods}

The AP is the definitive source for parliamentary transcripts of the Revolution, reconstructed from primary sources including transcripts, minutes, and newspaper reports. The French Revolution Digital Archive (\url{https://frda.stanford.edu}) is a digital version, with full-text speeches and TEI markup from the beginning of the NCA on 9 July 1789 through its end on 30 September 1791. After stop-word removal, each speech is represented as a count vector over vocabulary of $10,000$ most common words.  The resulting corpus contains $4,765,773$ words in $44,913$ speeches. Each speech is matched to a named speaker, and may be tagged with a ``role'': in some cases, a speech made by a delegate acting as the assembly president; in other cases, on behalf of a committee, either to introduce a report, or to comment on it. See Supplementary Information (SI; linked as ancillary file on the arXiv) for details on corpus preparation. All of the delegates to the Assembly, and all of the speakers in our data, were male.

We use LDA~\cite{Blei:2003tn} to model speeches as probability distributions over $K=100$ topics; the $j$th speech in the data is $s_{i}^{(j)}$, $1\leq i\leq K$. 

\subsection*{Novelty, Transience, Resonance}

\begin{figure}[h!]
\centering
\includegraphics[width=0.5\textwidth]{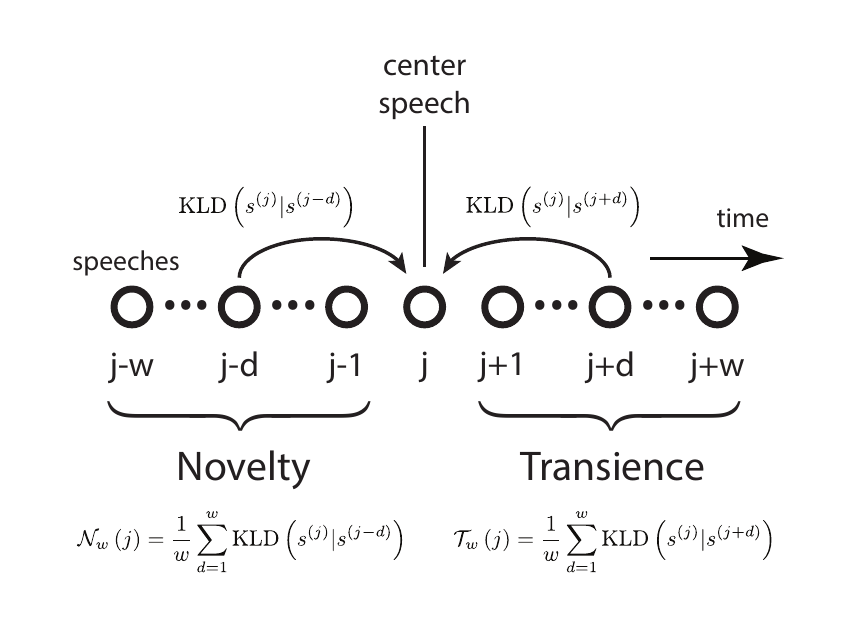}
\caption{Diagram of novelty and transience.  Each speech is represented by an LDA topic mixture.  The Kullback-Leibler Divergence $\mathrm{KLD}\left(s^{(j)}|s^{(j-d)}\right)$ measures surprise of speech at time $j$ given a speech $d$ steps away.  Novelty at scale $w$ is the mean surprise of a central speech given speeches in a past window of $w$ steps.  Transience is novelty under time reversal, measuring the mean surprise from the future.  A rhetorically effective speech shifts current conversation to a new issue with a differing vocabulary, manifesting itself as a surprise asymmetry where novelty exceeds transience.  We define this positive asymmetry as resonance (Eq.~\ref{eq:resonance}): the ability to break from the past and influence the future.  An illustration of this asymmetry is shown in the middle panel of Figure~\ref{fig:NTR_pedagogical}.}
\label{fig:innov_trans_illust}
\end{figure}

Novelty at the smallest timescales (a speech compared to the one just previous) is measured by the KLD of $s^{(j)}$ relative to $s^{(j-1)}$
\begin{align}
\mathrm{KLD}\left(s^{(j)}|s^{(j-1)}\right) = \sum_{i=1}^K s^{(j)}_i \log_2 \left ( \frac{s^{(j)}_i}{s^{(j-1)}_i} \right ).
\label{eq:stsKLD}
\end{align}
Averaging this measure further backwards in the debate allows us to consider longer-timescale trends, beyond the back-and-forth of a single exchange, to study the extent to which the current speaker has taken the debate in an unexpected direction given, say, the last twenty speeches. We refer to this quantity as novelty $\mathcal{N}$ at time $j$ on scale $w$,
\begin{align}
\mathcal{N}_w\left(j\right) = \frac{1}{w}\sum_{d=1}^w \mathrm{KLD}\left(s^{(j)}|s^{(j-d)}\right),
\label{eq:novelty}
\end{align}
illustrated in Fig.~\ref{fig:innov_trans_illust}. 

Any speech can break abruptly from its past---but the new patterns it introduces may not persist.  Consider an interjection that other speakers ignore in order to return to the matter at hand. That interjection would be surprising given its past, but equally surprising given its future, symmetric surprise under time reversal, because all the new information is lost, or transient.  In contrast, a rhetorically effective interjection might shift the conversation to a new issue with differing vocabulary.  This shift would appear as a surprise asymmetry about the interjection.  We define this asymmetry as ``resonance'', $\mathcal{R}$:
\begin{align}
 \mathcal{R}_w\left(j\right) &= \frac{1}{w}\sum_{d=1}^w \left [ \mathrm{KLD}\left(s^{(j)}|s^{(j-d)}\right) - 
 \mathrm{KLD}\left(s^{(j)}|s^{(j+d)}\right) \right ] \nonumber \\
 &\equiv \mathcal{N}_w\left(j\right) - \mathcal{T}_w\left(j\right)
\label{eq:resonance}
\end{align}
\noindent  Resonance is novelty minus transience, $\mathcal{T}$, where the latter is novelty in Eq.~\ref{eq:novelty} under time reversal. Novel speeches which also influence future discourse are pivot points in conversation. For a system in equilibrium, deviations are time-symmetric in the expectation---\emph{i.e.}, the measured novelty and transience at any point in time are, on average, equal. For a non-equilibrium system significant deviations may appear both for individual samples and for the overall average. Novelty's effectiveness, $\Gamma$, is the rate at which resonance increases with novelty, $\frac{\mathrm{d} \mathbf{E}[\mathcal{R}|\mathcal{N}]}{\mathrm{d} \mathcal{N}}$; approximated here using a linear model.


\section*{Acknowledgments}

\noindent J.H. acknowledges a Research Experience for Undergraduates National Science Foundation Grant \#ACI-1358567 at the Santa Fe Institute.



%
%
%
%
%



\end{document}